\documentclass[aps,twocolumn,showpacs]{revtex4}

\errorstopmode

\usepackage{graphicx}
\usepackage{amsmath}
\usepackage{amssymb}

\begin{document}

\title{Magnifying absolute instruments for optically homogeneous regions}

\author{Tom\'a\v{s} Tyc}
\affiliation{Institute of Theoretical Physics and Astrophysics,
Masaryk University, Kotl\'a\v rsk\'a 2, 61137 
Brno, Czech Republic}

\date{\today}
\begin{abstract}
We propose a class of magnifying absolute optical instruments with a positive
isotropic refractive index. They create magnified stigmatic images, either
virtual or real, of optically homogeneous three-dimensional spatial regions 
within geometrical optics.  \\

\end{abstract}
\maketitle

Most optical instruments, including a simple lens or sophisticated camera
lenses, have various types of aberrations. There exist, however, optical
instruments that are free of aberrations and provide sharp (stigmatic) images
of all points in some 3D region of space within geometrical optics; such
devices are called absolute instruments~\cite{BornWolf}.  A prototype of an
absolute instrument is Maxwell's fish eye, a device designed in 1854 by
J. C. Maxwell~\cite{MFE}. It uses positive refractive index and images
stigmatically the whole space.  Another type of absolute instrument is based on
materials with a negative refractive index. It was proposed by J. Pendry in
2000~\cite{Pendry2000} and later realized
experimentally~\cite{Fang2005-neg_n-silver_imaging}. Remarkably, both Maxwell's
fish eye and Pendry's lens are not limited by diffraction and provide
sub-wavelength resolution~\cite{Ulf2009-fisheye,Ulf2010-fisheye,Ma2011,
  Pendry2000, Fang2005-neg_n-silver_imaging}.

Among absolute instruments there is a class of a particular interest, namely
devices whose object and image spaces are optically homogeneous regions, i.e.,
regions with a uniform refractive index. Until recently, the only known such
devices were plane mirrors~\cite{BornWolf}.  This has been changed by a recent
excellent work of J. C. Mi\~nano~\cite{Minano2006} who proposed several new
absolute instruments imaging homogeneous regions and also showed that some
well-known optical devices such as Eaton lens or Luneburg
lens~\cite{Luneburg1964} are in fact absolute instruments as well. All of these
devices have unit magnification, giving an image of the same size as the
original object, and no magnifying absolute instrument for homogeneous regions
has been known.  We proposed a magnifying absolute instrument
recently~\cite{Tyc2010} based on a numerically found refractive index with
certain special properties, but our colleague Klaus Bering has later shown
analytically that such an index in fact does not exist~\cite{Tyc-topublish}.

Here we present several magnifying absolute instruments that provide stigmatic
images of homogeneous regions of 3D space with an arbitrary magnification. They
are all based on the same idea and provide either real or virtual images.  This
is the first proposal of a magnifying absolute instrument for homogeneous
regions that employs isotropic materials with positive refractive index.  We
will explain our idea first on a particular example of a magnifying absolute
instrument resembling Eaton lens~\cite{Eaton1952}, and then proceed to other
devices.

\begin{figure}
\begin{center}
\includegraphics[width=7cm]{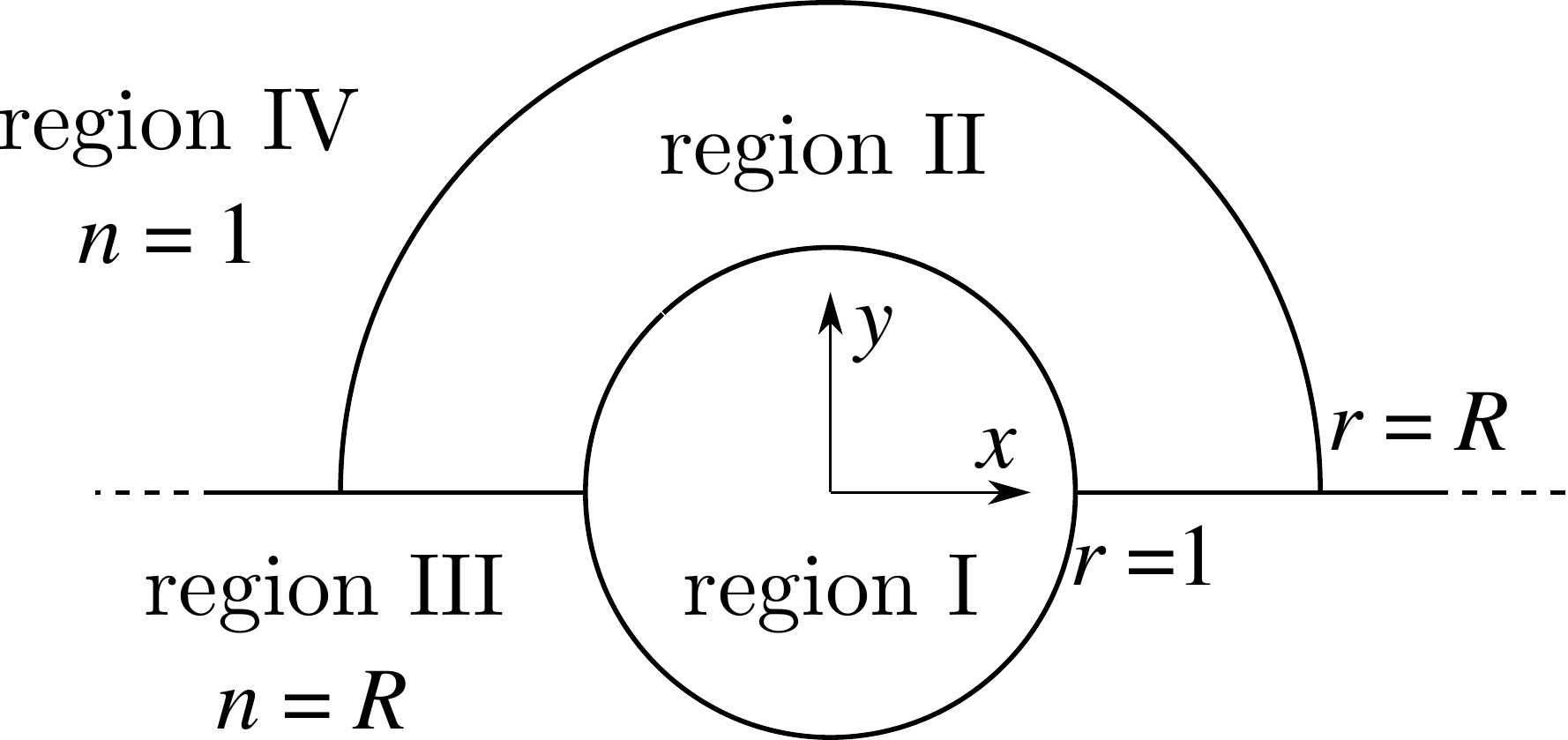}
\end{center}
\caption{Regions in the magnifying device based on Eaton lens. Regions III and
  IV that lie outside the device are the optically homogeneous object and image
  space.}
\label{regions}
\end{figure}

Our device consists of two distinct regions (see Fig.~\ref{regions}). The first
region (we will call it region I) is a sphere of unit radius, the second region
(II) occupies the space between two hemispheres with the radii 1 and $R>1$,
respectively, lying in the half-space $y>0$. Both regions are filled with a
spherically symmetric refractive index that we will denote by $n_{\rm I}(r)$
and $n_{\rm II}(r)$, respectively. The indices are chosen such that $n_{\rm
  I}(1)=n_{\rm II}(1)=R$ and $n_{\rm II}(R)=1$. The medium surrounding the lens
is composed of two parts as well. In the region $r\ge1,y<0$ (region III) the
refractive index is equal to $R$ while in the region $r\ge R,y>0$ (region IV)
the refractive index is equal to unity. Thus the index of the lens matches that
of the surrounding medium, apart from the annulus $1<r<R,y=0$, and also the
indexes at the border between regions I and II match each other.

\begin{figure}
\begin{center}
\includegraphics[width=7cm]{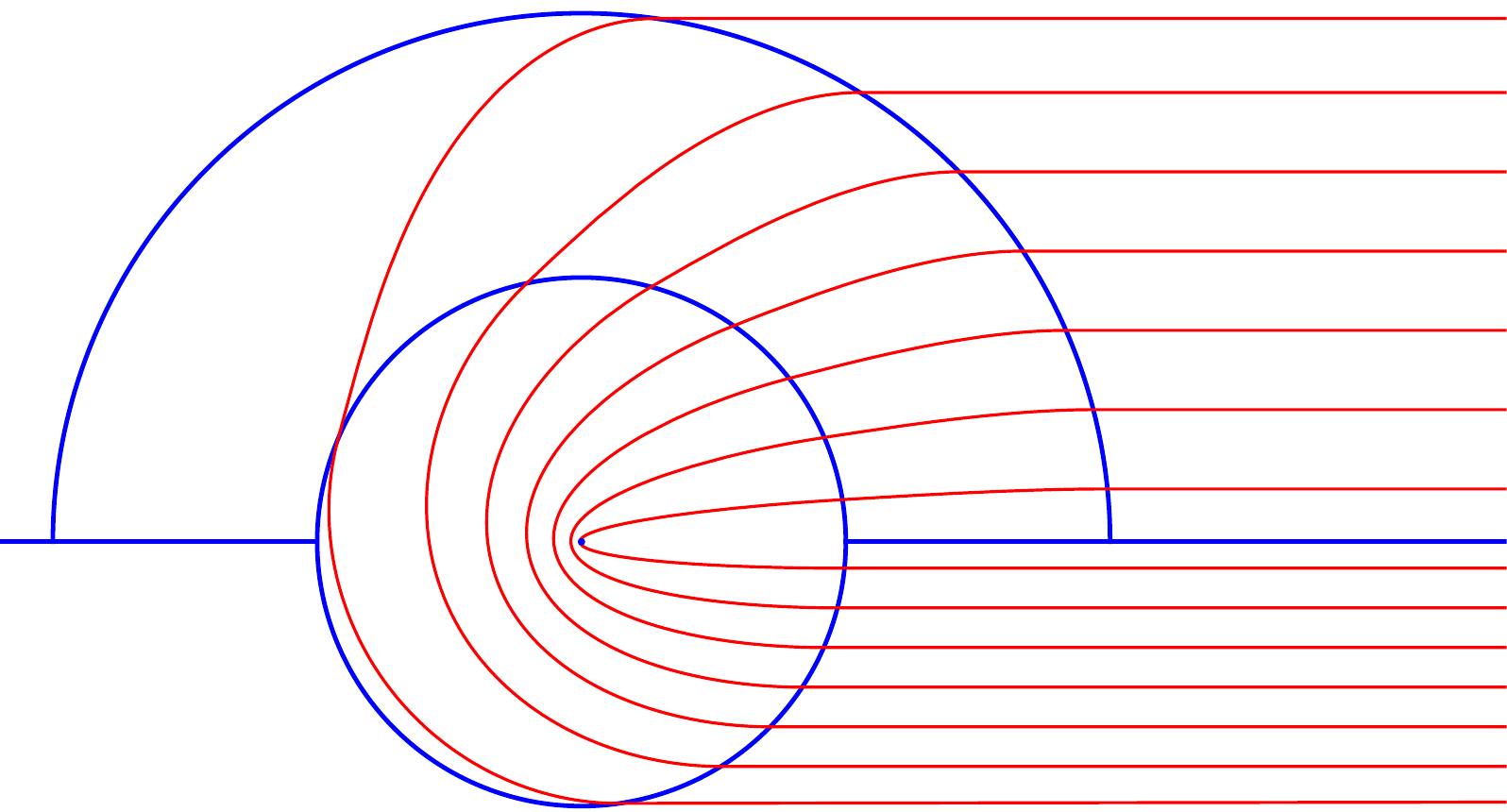}
\end{center}
\caption{Magnifying Eaton lens with $R=2$. Light rays incident on it from
  region IV are changed into rays in region III propagating in the opposite
  direction, with impact parameters reduced by the factor of $R$.}
\label{mag-eaton}
\end{figure}

The refractive index $n_{\rm II}(r)$ is chosen such that a light ray incident
from region IV to region II is bent towards the center, eventually crossing the
border between regions II and I. There is a large variety of refractive index
profiles that achieve this, one option is to choose $n_{\rm
  II}(r)=[1+c(r-1)(R-r)]R/r$ with $c>0$ sufficiently large, which we have also
used in our ray tracing simulations with $c=1$. We then design the refractive
index $n_{\rm I}(r)$ in region I such that the light ray coming from region II
is bent further and leaves region I for region III in exactly opposite
direction than was the original direction of the ray in region IV, see
Fig.~\ref{mag-eaton}. The performance of the device is thus similar to the
performance of Eaton lens; we can therefore call it ``magnifying Eaton
lens''. As we shall see, the difference is that the impact parameter of the
outgoing ray is $R$ times smaller than the impact parameter of the incoming
ray, and this fact is responsible for the lens magnification.

To design the refractive index $n_{\rm I}(r)$, we employ the standard method
for solving the inversion problem~\cite{Luneburg1964, Hendi2006}. We will
characterize rays in the lens by the quantity $L=nr\sin\alpha$ analogous to
mechanical angular momentum, where $\alpha$ is the angle between the radius
vector and the ray~\cite{Ulf-Thomas-book}. Angular momentum is conserved and
motion of a particle is planar in central potentials, and the same holds for
light ray in a spherically symmetric refractive index. Consider a ray
propagating in the plane $xy$ horizontally (i.e., in the direction of $x$-axis)
in region IV and entering region II of the lens at point A.  Since $n=1$ in
region III, the angular momentum $L$ is equal to the impact parameter of the
incoming ray. The polar angle $h$ swept by the ray in region II before entering
region I (say, at point B) can be calculated by the
expression~\cite{Landau-shorter}
\begin{equation}
  h(L)=L\int_1^{R}\frac{{\rm d} r}{r\sqrt{[rn_{\rm II}(r)]^2-L^2}}.
\label{deltaphi}
\end{equation}
$h(L)$ is at the same time the change of the ray direction during propagation
in region II from point A to B. This is because the product $nr$ is the same at
both points A and B, so is $L=nr\sin\alpha$, and therefore the angle $\alpha$
between the ray and radius vector is the same at B as is in A.

\begin{figure}
\begin{center}
\includegraphics[width=5cm]{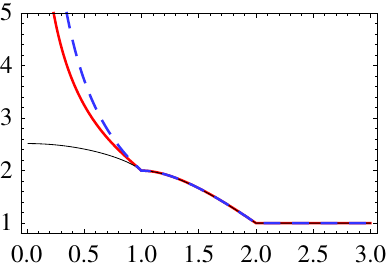}
\end{center}
\caption{Refractive index in regions I and II of magnifying Eaton lens (thick
  red), Luneburg lens (thin black) and magnifying invisible sphere (dashed
  blue) with $R=2$.}
\label{mag-luneburg-index}
\end{figure}

The scattering angle $\chi$ (change of ray direction) corresponding to motion
in region I must therefore be
\begin{equation}
\chi(L)=\pi-h(L)\,,
\end{equation} 
which ensures that the total change of ray direction during motion in regions II
and I is $\pi$. Solving the inversion problem, we then arrive at the following
implicit equation for the refractive index $n_{\rm I}(r)$~\cite{Luneburg1964,
  Hendi2006}:
\begin{equation}
  n_{\rm I}(r)=R\exp\left[\frac1\pi\int_{rn_{\rm I}(r)}^{R}\frac{\chi(L){\rm d}
      L}{\sqrt{L^2-[rn_{\rm I}(r)]^2}}\right] \,.
\label{index}
\end{equation}
This way the refractive index is expressed analytically, although not
explicitly. The refractive index in regions I and II is shown in
Fig.~\ref{mag-luneburg-index}.

Now we have to show that the device we have just described is indeed an
absolute magnifying instrument, i.e., it provides stigmatic image of some 3D
region of space.  First we note that from conservation of $L$ it follows that
the impact parameter of the outgoing ray in region II is $R$-times smaller than
the impact parameter of the incoming ray in region IV.  Second we note that
although we considered a horizontal ray in the $xy$-plane in our construction,
the lens will have a similar effect on most other rays too, i.e., rays incident
on it from region IV will be changed to rays moving in the opposite direction
in region III and having $R$-times smaller impact parameters. This is caused by
the spherical symmetry of the refractive index in regions I and II. There will
also be rays for which this does not happen, namely the ones that at some point
cross the interface between regions II and III, but still for an infinite
number of rays the lens does the job it is designed for.
 
\begin{figure}
\begin{center}
\includegraphics[width=8cm]{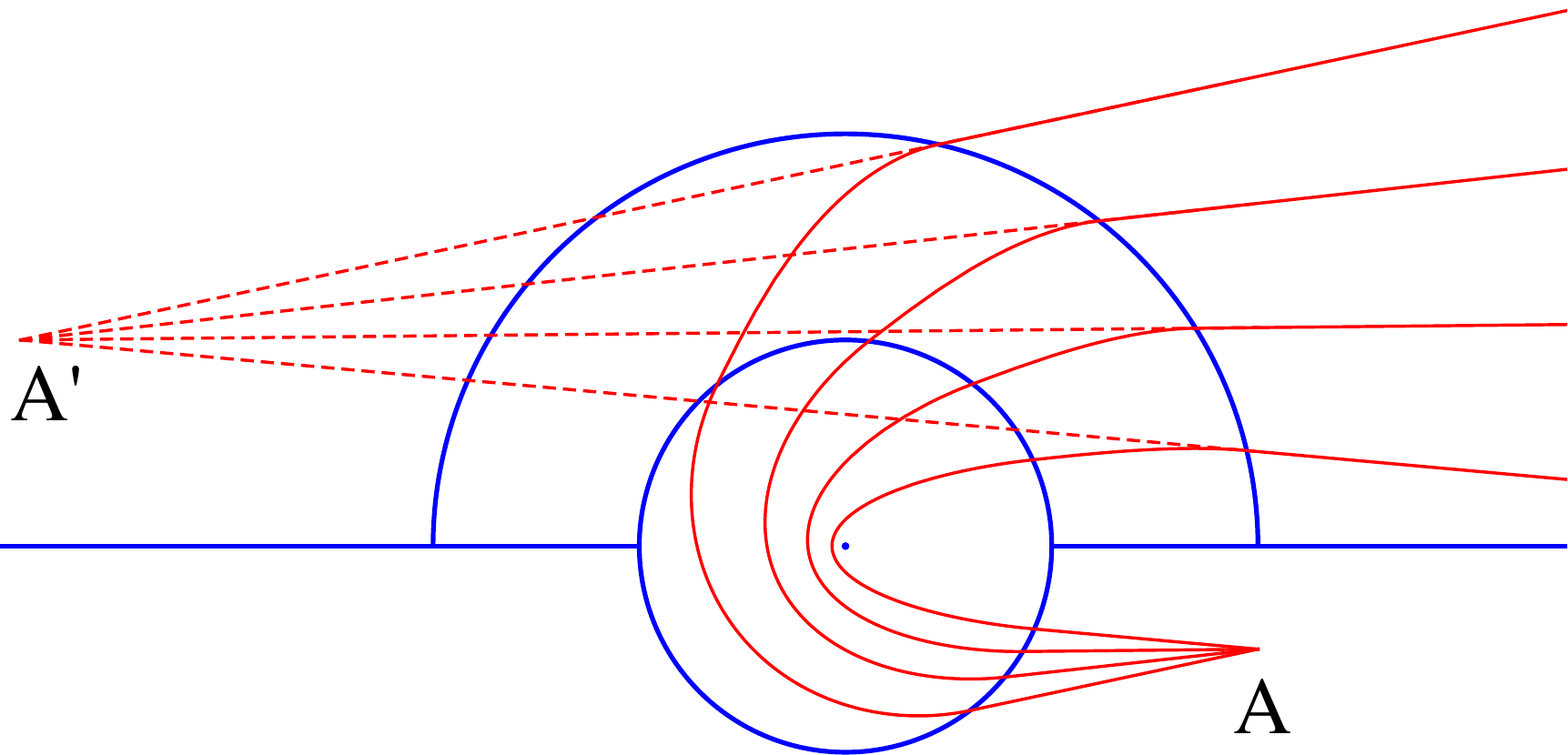}
\end{center}
\caption{Image formation in magnifying Eaton lens. Light rays emerging from a
  point A at $\vec r_{\rm A}$ in region III, after passing through the lens,
  propagate in region IV as if they originated from a point ${\rm A'}$ at
 $\vec r_{\rm A'}=-R\vec r_{\rm A}$. The image is virtual and is magnified
  $R$-times.}
\label{mag-eaton-imaging}
\end{figure}

Now consider a collection of rays emerging from some point A at radius vector
$\vec r_{\rm A}$ in region III and incident on region I of the lens.  As we
have seen, these rays will be transformed by the lens into rays propagating in
region IV, each parallel to the original ray in region III.  Therefore the
lines on which these outgoing rays lie intersect at the point A$'$ with radius
vector $\vec r_{\rm A'}=-R\vec r_{\rm A}$, which this way becomes the virtual
image of the point A (see Fig.~\ref{mag-eaton-imaging}), and the magnification
of the device is clearly equal to $R$.

\begin{figure}
\begin{center}
\includegraphics[width=7cm]{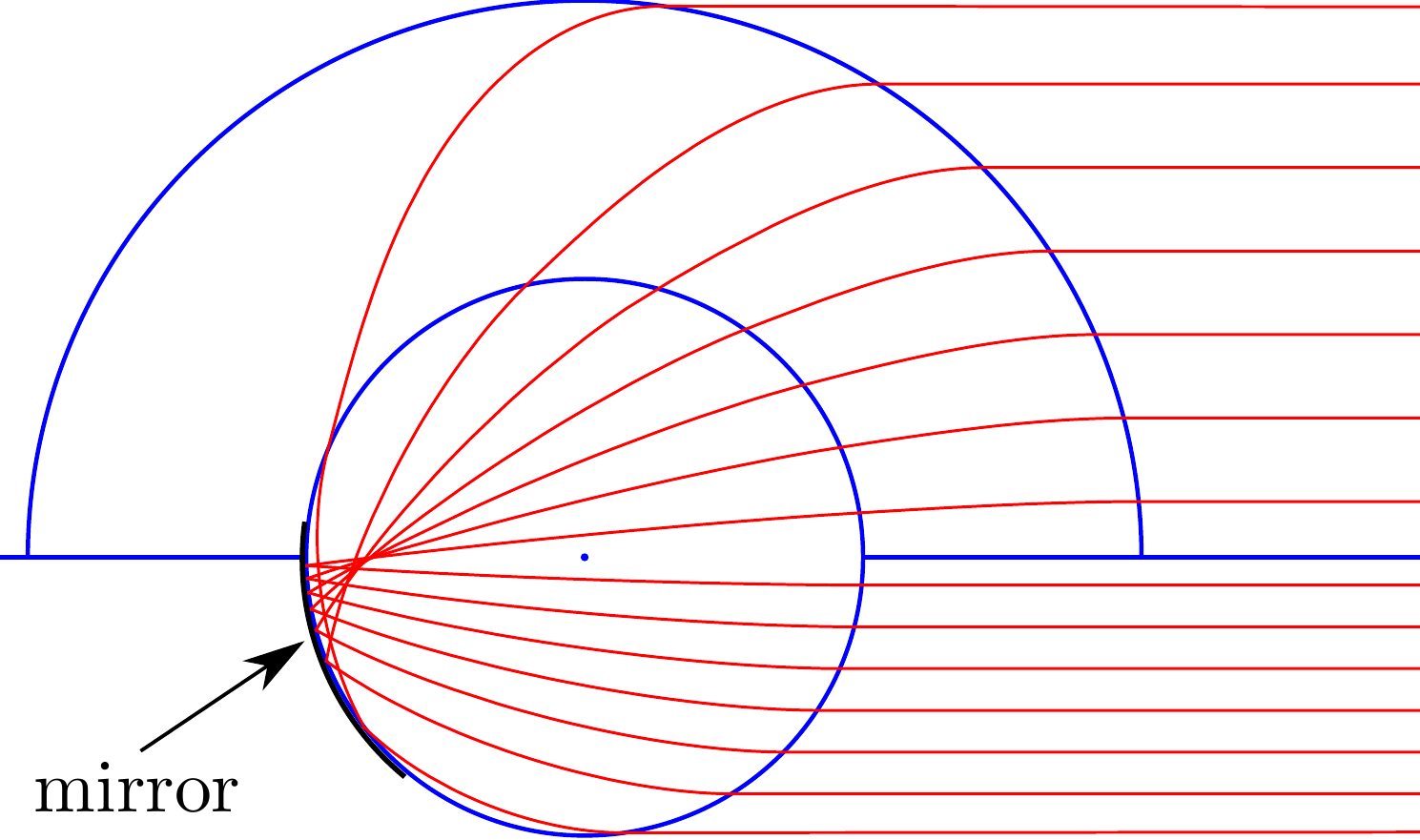}
\end{center}
\caption{Magnifying Luneburg lens with $R=2$. A spherical mirror covers part
  of the interface between regions I and II. The effect on rays coming from
  region IV is the same as that of magnifying Luneburg lens and therefore it
  also provides a magnified virtual image.}
\label{mag-luneburg}
\end{figure}

As can be shown, the refractive index in region I diverges for $r\to0$. To
avoid this singularity, we can modify the lens by utilizing the fact that
Luneburg lens~\cite{Luneburg1964} equipped with a spherical mirror on its
surface has the same effect on the incoming rays~\cite{Minano2006} as Eaton
lens.  Imagine we place a mirror on the part of the interface between regions I
and II, allowing the rays in region I to be reflected before re-entering region
II. We again require that the outgoing rays move in the opposite direction with
respect to the incoming rays, but now rays with small impact parameters do not
have to make a rapid turn near the origin at it was the case with magnifying
Eaton lens, and therefore the index will not need a singularity there. It can
be shown by simple geometrical considerations that in this case the required
scattering angle in region I corresponding to the ray segment between point of
entrance to region I and the point of incidence on the mirror is
\begin{equation}
\chi(L)=\arcsin\frac LR-\frac{h(L)}2\,,
\end{equation} 
which, after substitution to Eq.~(\ref{index}) leads to the refractive index
shown in Fig.~\ref{mag-luneburg-index} for $R=2$. Ray tracing in the
magnifying Luneburg lens is shown in Fig.~\ref{mag-luneburg}.

\begin{figure}
\begin{center}
\includegraphics[width=7cm]{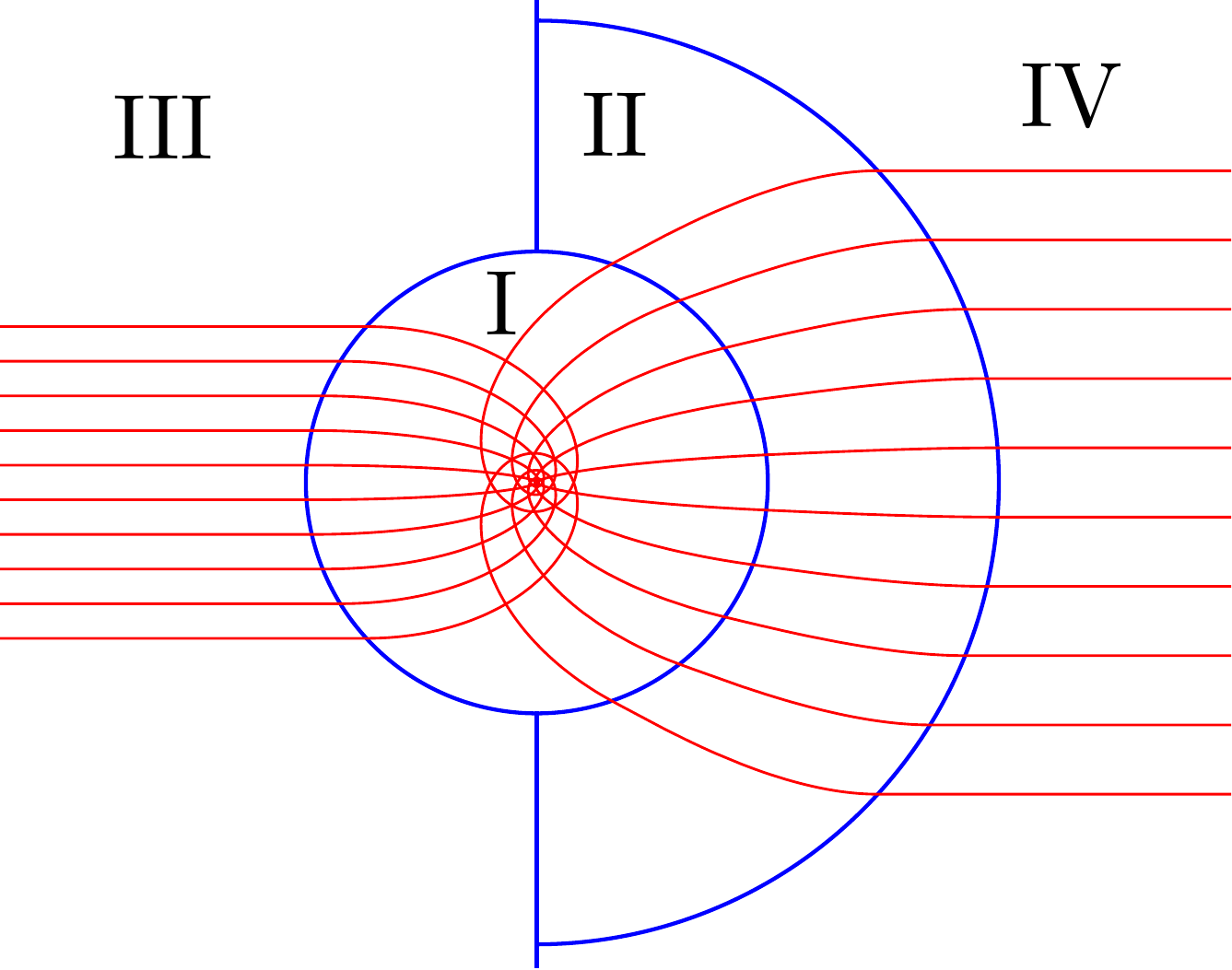} 
\end{center}
\caption{Magnifying device derived from the invisible sphere~\cite{Hendi2006}.
  The rays make loops in region I and after leaving the device, they propagate
  in the original direction. This device again forms a virtual image.}
\label{invisible}
\end{figure}

Another interesting magnifying device can be derived by our method from the
invisible sphere described in~\cite{Hendi2006}.  Here we require again that the
rays leaving the lens propagate parallel to their original direction, but this
time go forwards instead of backwards.  The scattering angle in this case is
$\chi(L)=2\pi-h(L)$. If the lens should work well for rays with the direction
close to $x$-axis, the border between regions III and IV now has to be in the
$yz$ plane and similarly region II now lies at $x>0$ instead of $y>0$. The
virtual image of a point A is now formed at point A$'$ with $\vec r_{\rm
  A'}=R\vec r_{\rm A}$. Ray tracing in this lens is shown in
Fig.~\ref{invisible}, the refractive index is in Fig.~\ref{mag-luneburg-index}.

\begin{figure}
\begin{center}
\includegraphics[width=6cm]{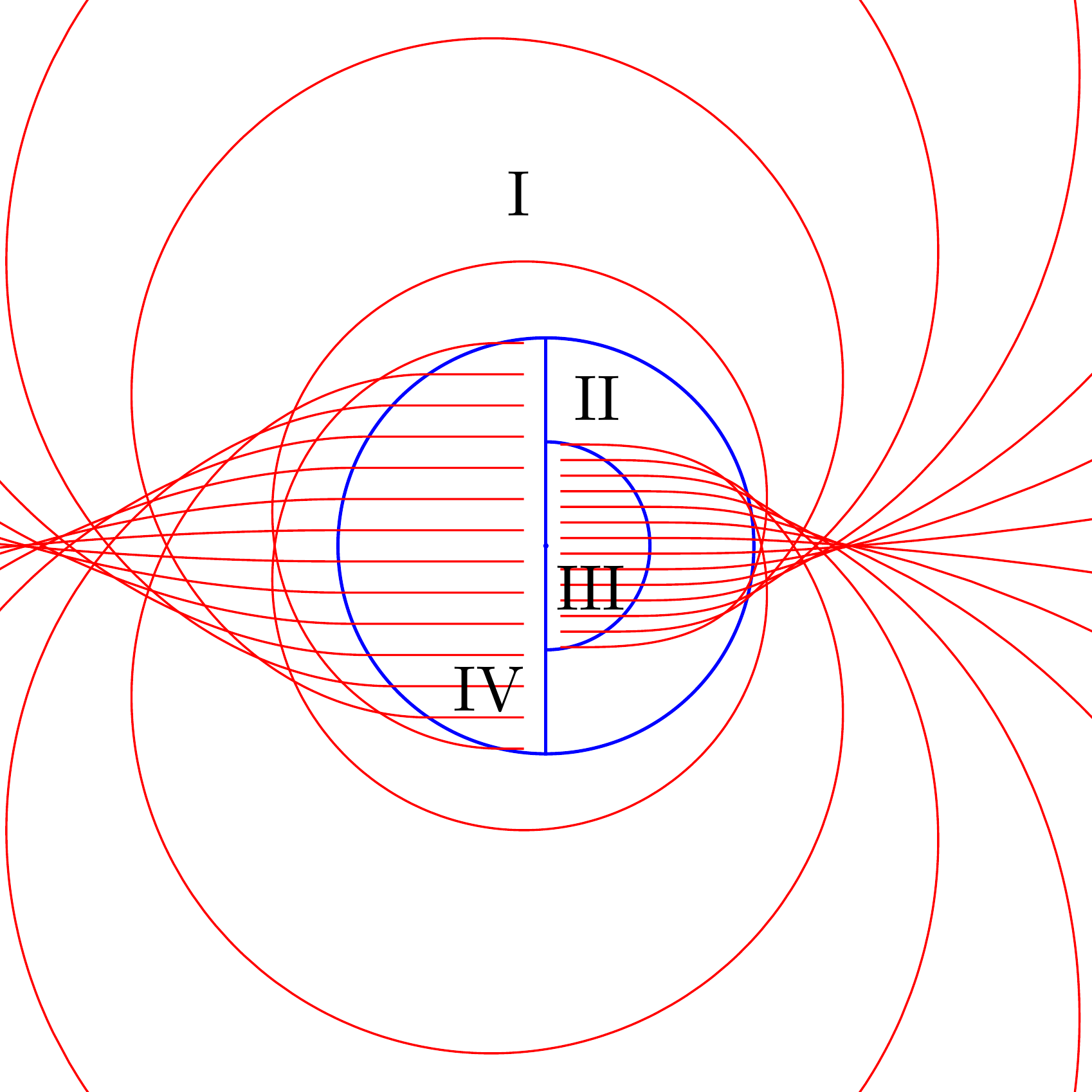}
\end{center}
\caption{Inside-out magnifying instrument with the regions marked. The rays
  enter the lens from region III inside the lens (the object space) and after
  making loops in region I, which now extends to infinity, enter region IV
  (image space) in their original direction. }
\label{abs1}
\end{figure}

The last device we will discuss is a magnifying absolute instrument that
provides real images. Now we have to arrange regions I -- IV in a somewhat
different fashion, see Fig.~\ref{abs1}: region I is given by the condition
$r>R$, so it is the whole 3D space with the exception of the sphere of radius
$R$.  Region II occupies the space between two hemispheres with the radii 1 and
$R>1$, respectively, lying in the half-space $x>0$. Region III is the unit
hemisphere at $x>0$ and region IV is a hemisphere of radius $R$ at
$x<0$. Refractive indices in regions II, III and IV are as before.  Light rays
now enter the lens (region II of it) from region III, i.e., from the
inside. After having been bent in region II, they propagate in region I and we
require that when they enter region IV, they are parallel to their original
direction in region III, see Fig.~\ref{abs1}. To find the refractive index
$n_{\rm I}(r)$ that achieves this, we have to solve an ``outer'' inversion
problem instead of the usual ``inner'' problem. This can be done by employing
inversion in the sphere of radius $R$ which transforms the outer problem to the
inner one. A careful analysis of the scattering angle $\chi'$ in the
transformed  problem reveals that $\chi'(L)=4\arcsin(L/R)-h(L)$, which
then gives the transformed refractive index $n'_{\rm I}(r')$ as a function of
the transformed radius $r'=R^2/r$ with the help of formula~(\ref{index}), but
with the ``$R$'' omitted in front of the exponential since we require $n_{\rm
  I}'(R)=1$. The index $n_{\rm I}(r)$ can then be calculated as
\begin{equation}
  n_{\rm I}(r)=\frac{R^2}{r^2}\,n'_{\rm I}(R^2/r)\,,
\end{equation} 
which follows from the equality of optical paths in the original and
transformed region I.

\begin{figure}
\begin{center}
\includegraphics[width=5.7cm]{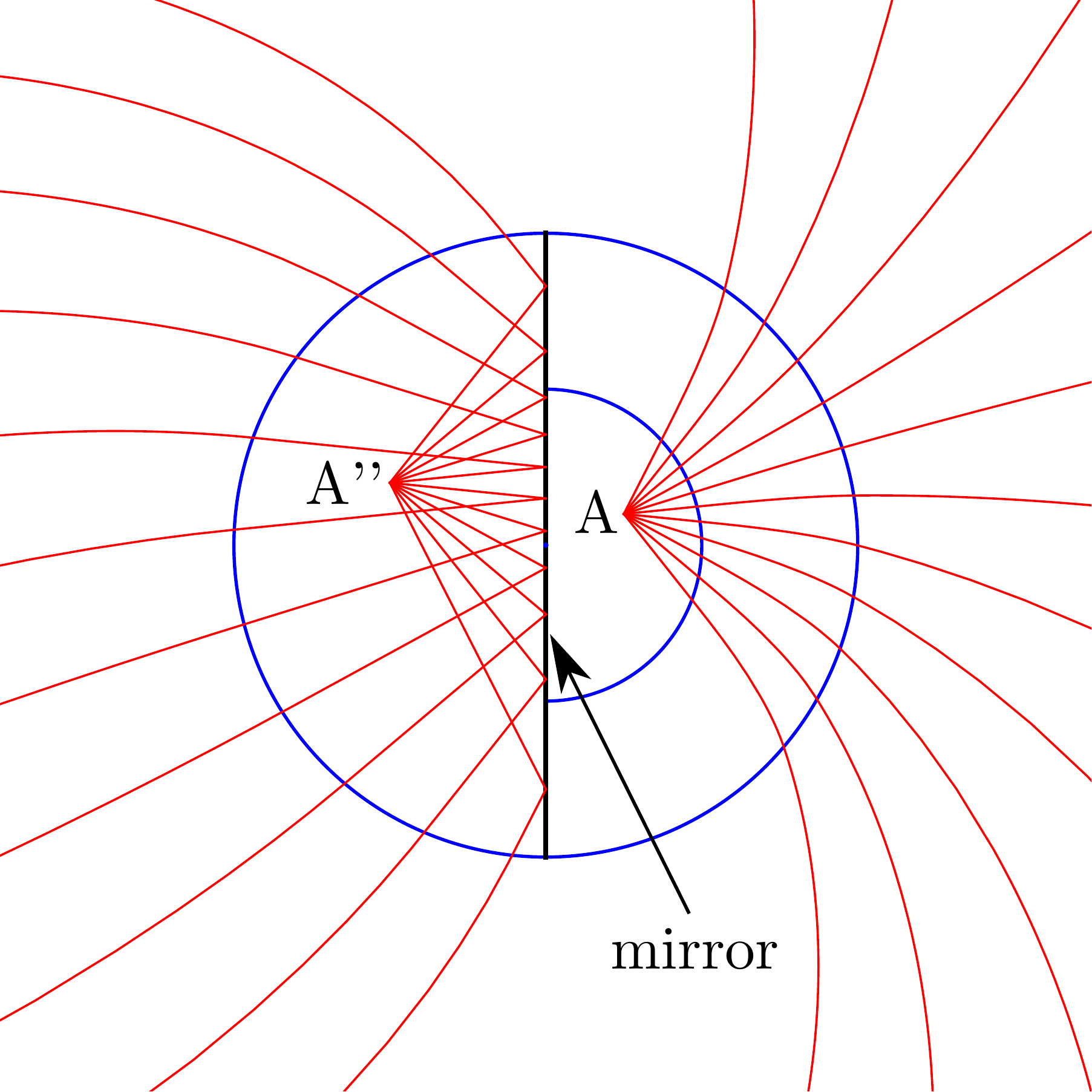}
\end{center}
\caption{Image formation on the instrument from Fig.~\ref{abs1}. Rays
  originating at point A in region III and converging in region IV to a point
  outside this region are reflected by the mirror to the real image A$''$ of
  A.}
\label{abs2}
\end{figure}

To see that this device indeed creates a real magnified image, consider rays
emerging from some point A at radius vector $\vec r_{\rm A}$ in region III, see
Fig.~\ref{abs2}.  The rays get to region IV assuming their original direction
and head towards the point A$'$ at $\vec r_{\rm A'}=R\vec r_{\rm A}$.  This
point lies outside of region IV and therefore the image is virtual. However, we
can take advantage of the fact that the rays are converging and place a mirror
at the flat interface of region IV, see Fig.~\ref{abs2}. This way the virtual
image it turned into a real image at point A$''$, which is a mirror image of
A$'$ in the plane $x=0$. Making the mirror double-sided, also rays emerging
from the point A to the left will contribute to forming the image.

In conclusion, we have proposed several absolute optical instruments that
create magnified stigmatic images of homogeneous 3D region. They are all
designed by the same general idea. Especially appealing is the lens giving a
real magnified image and the magnifying Luneburg lens with its moderate
refractive index range; for $R$ not too large, it should be possible to realize
the latter for near infrared or even visible light using e.g. graded index
structures in silicon~\cite{Gabrielli} or diamond.  Further research will
reveal whether some of these devices, e.g. the lens giving the real image,
could provide sub-wavelength resolution similarly to Maxwell's fish
eye~\cite{Ulf2009-fisheye,Ulf2010-fisheye, Ma2011}.  Magnifying absolute
instruments could find their applications in various fields, for example in
photolitography, but more importantly, our research has shown that such devices
exist at all, something that was not clear until this
date~\cite{BornWolf,Minano2006}.

This work was supported by grants MSM0021622409 and MSM0021622419.

\end{document}